# In-situ study and modeling of the reaction kinetics during molecular beam epitaxy of GeO$_2$ and its etching by Ge

Wenshan Chen, Kingsley Egbo, Hans Tornatzky, Manfred Ramsteiner, Markus R. Wagner, Oliver Bierwagen

*Paul-Drude-Institut für Festkörperelektronik, Leibniz-Institut im Forschungsverbund Berlin e.V., Hausvogteiplatz 5–7, 10117 Berlin, Germany*

## Abstract

Rutile GeO$_2$ has been predicted to be an ultra-wide bandgap semiconductor suitable for future power electronics devices while quartz-like GeO$_2$ shows piezoelectric properties. To explore these crystalline phases for application and fundamental materials investigations, molecular beam epitaxy (MBE) is a well-suited thin film growth technique. In this study, we investigate the reaction kinetics of GeO$_2$ during plasma-assisted MBE using elemental Ge and plasma-activated oxygen fluxes. The growth rate as a function of oxygen flux is measured *in-situ* by laser reflectometry at different growth temperatures. A flux of the suboxide GeO desorbing off the growth surface is identified and quantified *in-situ* by the line-of-sight quadrupole mass spectrometry. Our measurements reveal that the suboxide formation and desorption limits the growth rate under metal-rich or high temperature growth conditions, and leads to etching of the grown GeO$_2$ layer under Ge flux in the absence of oxygen. The quantitative results fit the sub-compound mediated reaction model, indicating the intermediate formation of the suboxide at the growth front. This model is further utilized to delineate the GeO$_2$-growth window in terms of oxygen-flux and substrate temperature. Our study can serve as a guidance for the thin film synthesis of GeO$_2$ and defect-free mesa etching in future GeO$_2$-device processing.

**Introduction**

In contrast to wide-bandgap semiconductors, such as SiC or GaN, ultra-wide-bandgap (UWBG) semiconductors with a bandgap larger than 3.4 eV show great advantages for power electronics as their bandgap allows for a larger breakdown field strength that enables much higher power density and thus make practical contributions for energy saving[1–4]. Such merits offer UWBG materials a competitive edge over traditional narrow band gap semiconductors. So far, β-$Ga_2O_3$, diamond, and AlN/AlGaN are the mainstream UWBGs in the fundamental research of high-power devices and a series of impressive progress has been achieved[5–10]. However, β-$Ga_2O_3$ encounters insurmountable bottlenecks such as poor thermal conductivity and doping asymmetry which limit its further breakthrough, while diamond and AlN/AlGaN are hampered by the high AlN and diamond substrate price and the difficulties in homogeneous epitaxy[8,11]. Hence, alternative UWBG materials are indispensable to meet the demand of next generation power devices.

Rutile $GeO_2$ (r-$GeO_2$) is emerging as promising UWBG oxide material as it possesses a bandgap of ≈4.7 eV[12] and exhibits superior thermal conductivity[13] as well as higher predicted carrier mobility compared to β-$Ga_2O_3$.[14] More encouragingly, $GeO_2$ has also been predicted to exhibit a bipolar doping behavior, a property that is very rare in most oxide semiconductor materials[15]. The ability to dope r-$GeO_2$ both n- and p-type can thus promote desirable oxide homojunction applications. These intrinsic prominent features of r-$GeO_2$ highlight its tremendous prospects in efficient high voltage high power electronic applications.[14] Besides the rutile phase, $GeO_2$ can also form competing amorphous glass[16] and piezoelectric quartz-like phases,[17] necessitating a stabilization of the desired phase by epitaxy.

To date, several efforts towards growth of single crystalline r-$GeO_2$ using pulsed laser deposition, mist-chemical vapor deposition (mist-CVD) and MBE have been reported and recently $GeO_2$ has also achieved the transition from the amorphous phase to the polycrystalline one by post-annealing [18–23]. Among these techniques, MBE is well suited for development of r-$GeO_2$ following its success in pioneering the development of $Ga_2O_3$ due to the possibility of controlled epitaxial growth and doping using the most simple chemistry towards high quality films.[10] A comprehensive understanding of the reactions during MBE of $GeO_2$ and the rational establishment of a growth window are essential for the controlled synthesis of high-quality thin films. Specific research on this growth kinetics is lacking at present. Solely, the seminal work by Chae *et al.* [24] demonstrated the MBE growth of r-$GeO_2$ using a source of the suboxide GeO in ozone-MBE. They related the observed absence of growth at high growth temperature or low ozone partial pressure to a full desorption of the supplied GeO.

In metal-oxide MBE, $Me_xO_y$ is usually grown by the oxidation of an evaporated metal (Me) or suboxide flux by a flux of molecular oxygen, plasma-activated oxygen, or ozone on a heated substrate in vacuum. This process has been well studied in several oxide materials. The growth kinetics during this process mediated by the suboxide phase as intermediate reaction product has been studied in depth for $Ga_2O_3$, $In_2O_3$, and $SnO_2$ grown from the metal source [25–30] and for $Ga_2O_3$ grown from the suboxide source.[31]

In this letter, we extend this understanding to explore the growth process of GeO₂ without focusing on the formed phase. We present a detailed, systematic study investigating the formation of oxide and volatile suboxide during MBE growth of GeO₂ by monitoring *in-situ* the growth rate ($\Gamma$) or etch rate and desorbing species from the growth surface. We observed, that the suboxide formation and its desorption reduced the growth rate under oxygen-deficient or high temperature growth conditions, and resulted in etching of the grown GeO₂ layer by a pure Ge-flux. Based on the two-step oxidization mechanism that describes the MBE growth of binary oxides such as In₂O₃, Ga₂O₃ and SnO₂[32], we empirically modeled the GeO₂ growth via the intermediate formation of its suboxide GeO. Our measured growth rate as a function of growth temperature ($T_G$) and oxygen flux ($\Phi_O$) is in good quantitative agreement with this model. The growth rate of GeO₂ was found to be governed by the competition between GeO₂ accumulation and GeO desorption as a function of growth parameters.

**Experimental details**
For this study, GeO₂ layers were grown on 2-inch diameter, single-side polished, c-plane sapphire (Al₂O₃(0001)) wafers by plasma-assisted MBE (PAMBE). The rough backside of the substrate was sputter-coated with titanium to allow for non-contact substrate heating by radiation from the substrate heater. The growth temperature was measured with a thermocouple placed between the heater and the substrate. A standard shuttered effusion cell was used to evaporate Ge (7N purity) from a BN crucible. To avoid the potential cracking of the cell crucible during solidification of Ge, the Ge cell was ramped down slowly across the melting point. A radio frequency plasma source with a mass flow controller supplied activated oxygen from the research-grade O₂ gas (6N purity). The beam equivalent pressure (BEP), proportional to the particle flux, was measured by a nude filament ion gauge positioned at the substrate location. A Ge cell temperature of 1300°C was maintained throughout the experiments and the corresponding BEPs as function of cell temperature are shown in Fig S1. The BEPs are given in units of mbar and are converted into the equivalent particle flux (atoms nm⁻²s⁻¹) by multiplying the measured growth rate of the GeO₂ layer under conditions of full Ge incorporation by the cation number density[33]. A value of 2.4x10²² cm⁻³ for rutile GeO₂ is used, and the full Ge incorporation is ensured by growing in the oxygen-rich regime and at low substrate temperature of 300 °C where GeO or Ge desorption are negligible, a fixed Ge flux of $\Phi_{Ge}$= 4.06 nm⁻²s⁻¹ was obtained based on the measured growth rate of 1.7 Å/s (corresponding to 612 nm/h). The oxygen-rich regime was identified by measuring the growth rate as function of oxygen flux and will be discussed below. The radio frequency power of the plasma source was maintained at 300 W and the oxygen mass flow was set to 0.5, 0.8, 1.0, 1.5 and 2.0, standard cubic centimeters per minute (sccm), respectively. These oxygen flows were converted to active oxygen fluxes (in atoms nm⁻²s⁻¹) by multiplying the growth rate at stochiometric flux conditions with anion number density as described in Ref.[33]. In a separate experiment, we attempted to use molecular oxygen to oxidize Ge and found that the molecular oxygen was not reactive enough to form GeO₂ even at a high flow of 2 sccm.

*In-situ* laser reflectometry (LR) with a laser emitting at a wavelength of 650 nm at an incident angle of 30 degrees was used to measure the growth and etch rates for all layers. Using this laser, one oscillation period corresponds to a change of layer thickness by 200 nm taking into account the refractive index of our films of 1.6 as determined by spectroscopic ellipsometry. The flux of desorbing species during growth and etching was measured *in-situ* by line-of-sight quadrupole mass spectrometry (QMS). The QMS signal was converted into the equivalent growth rate by scaling it to reference conditions (high temperature of 800°C with 1 sccm plasma-activated oxygen flow) where the provided Ge flux completely desorbed as GeO. Further, growth was *in-situ* monitored by reflection high-energy electron diffraction (RHEED), which allowed us to corroborate our interpretation of the QMS and LR measurements.

The $GeO_2$ phase was identified ex-situ by Raman spectra acquired after growth using a Horiba LabRAM Evolution 800 mm Spectrograph with 473 nm laser excitation in backscattering geometry. The refractive index, required for the LR growth-rate-calibration, was determined by spectroscopic ellipsometry using a variable-angle spectroscopic ellipsometer based on a grating monochromator equipped with an auto retarder, in the range from 2500-250nm (~0.5-5eV). This measurement yields the amplitude ratio between the parallel and perpendicular component of the reflected polarized light from the sample with respect to the plane of incidence (Ψ) and the corresponding phase shift between these components (Δ). From this the refractive index *n* and the extinction coefficient *k* were determined, using a multi-layer model which contained a model refractive index $n(\hbar\omega)$ of the c-plane $Al_2O_3$ substrate layer[34] and the $GeO_2$ layer of interest and as top layer an effective medium approximated surface roughness layer based on a model by Bruggeman.[35] The refractive index was extracted based on a Cauchy model.[36]

**Results and discussion**

Figure 1(a) presents the reflected laser intensity measured by LR as a function of time during the $GeO_2$ growth and its subsequent etching by Ge. The two observed oscillatory periods, of growth (under Ge and activated oxygen flux, "(i)"—"(ii)") and etching (under Ge flux only, starting at (iii)), allow us to determine the growth and etch rate. Figure 1(b) shows the species desorbing off the substrate measured *in-situ* by QMS during a similar growth and etching experiment. Here the applied $T_G$ was 300°C while the oxygen flux was 8.12 $nm^{-2}s^{-1}$. Under this oxygen-rich and low $T_G$ condition, the provided Ge flux did not result in any appreciable desorbing cation or suboxide flux, implying the full incorporation of the Ge into the oxide film. In the process of etching, the substrate temperature ($T_{sub}$) was increased to 700°C to facilitate GeO desorption. While the film is stable at this temperature in vacuum (invariant LR signal in Figure 1 (a) and negligible QMS signal in Figure 1 (b) directly before (iii)), the impinging Ge atoms decompose the already grown $GeO_2$ layer, resulting in a reduction of its thickness (oscillation starting at (iii) in Figure 1 (a)) and a sharp increase of the GeO desorbing flux (seen in Figure 1 (b)). During etching, the detected Ge signal is an artifact of the fragmentation of GeO molecules by the electrons of the ionizer of the QMS. The grown $GeO_2$ layer could be etched down to the substrate, which is indicated by a disappearing GeO and an increased Ge desorbing flux in Figure 1(b)) and by reappearance of the streaky RHEED pattern of the sapphire substrate (Figure 2).

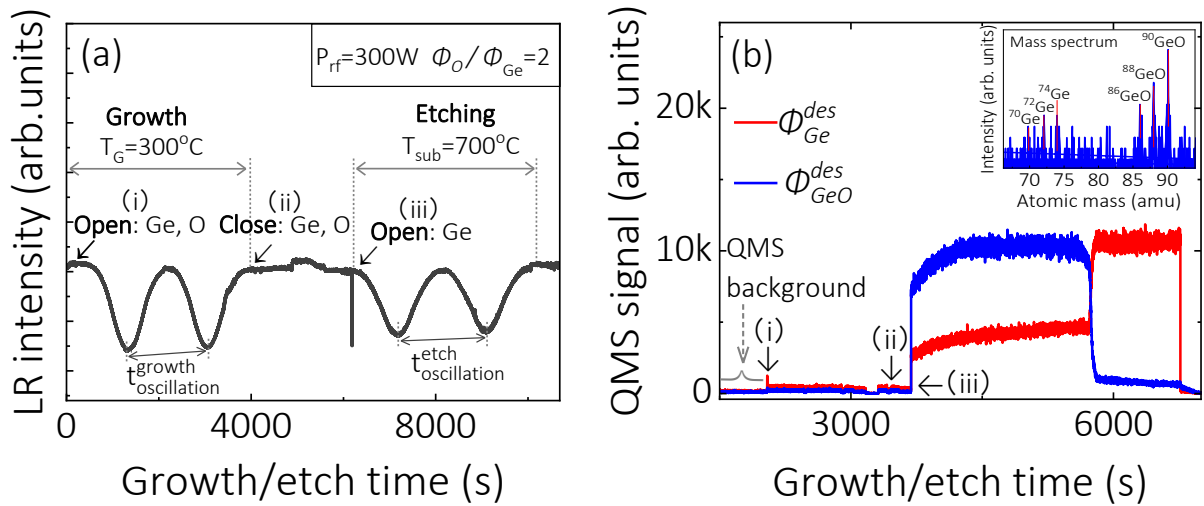

Figure 1. (a) LR signal during growth and etching of $GeO_2$ with corresponding oscillation periods indicated. The events of opening and closing Ge and O-fluxes are marked (i)—(iii). (b) The corresponding desorbing Ge and GeO fluxes measured by QMS in a similar growth and etching experiment. The events (i)—(iii) correspond to those in (a). The inset shows the mass spectrum of Ge and GeO detected by line-of-sight quadrupole mass spectroscopy (QMS). The Ge flux is fixed at $\Phi_{Ge}= 4.06$ $nm^{-2}s^{-1}$.

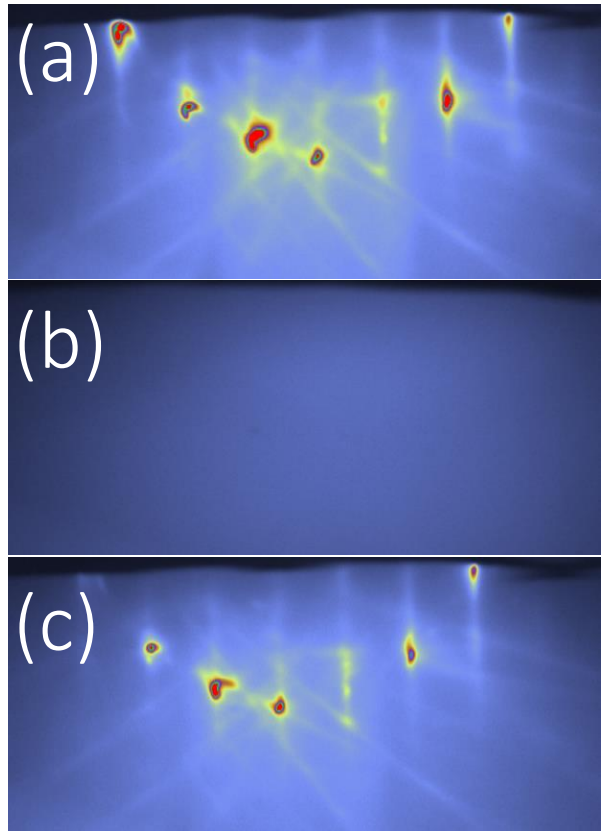

*Figure 2. Evolution of the RHEED pattern during growth and subsequent in-situ etching of GeO$_2$: (a) c-plane sapphire substrate after oxygen plasma cleaning before growth; (b) after growth of an amorphous GeO$_2$ layer at 300°C and (c) after subsequent etching of the layer at 700°C by Ge.*

The net reactions for the growth and etching are, Ge(g) + 2O(g) → GeO$_2$(s) and Ge (g) + GeO$_2$(s) → 2GeO(g), respectively. The gas and solid phases are labeled as g and s, respectively. Here, the same proportion of Ge is required to grow and etch the same amount of GeO$_2$ layer, therefore the etch rate was equal to the growth rate (cf. Figure 1(a)). These findings are in accordance with the PAMBE growth and etching of SnO$_2$ mediated by its suboxide SnO.[37]

The intensity and absence of diffraction spots or streaks in the RHEED image (Figure 2(b)) of the GeO$_2$ layer grown at 300°C indicates formation of the amorphous GeO$_2$ phase. The same holds true for a multi-layer sample grown at 500°C and 600°C, using several different O-fluxes at each temperature to extract the growth kinetics. Ex-situ Raman spectroscopy and spectroscopic ellipsometry confirmed the formation of GeO$_2$ in the amorphous phase (Figs. S2, S3) and allowed us to extract the refractive index of 1.6 at the wavelength of our LR (see Fig. S3).

The LR signal of this multilayer sample is shown in Figure 3.

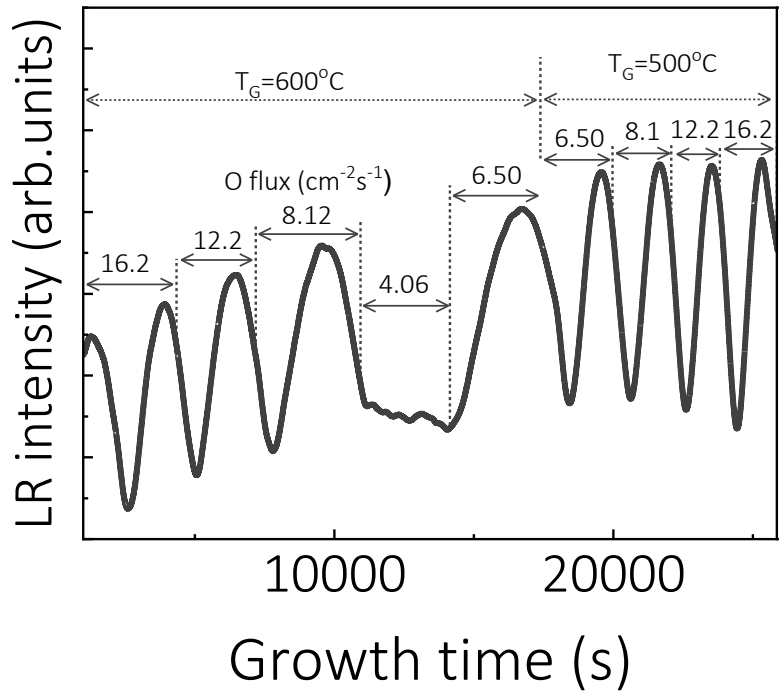

Figure 3. LR signal during GeO$_2$ growth with different growth conditions. The extracted growth rates are summarized in Figure 4. The Ge flux is fixed at Φ$_{Ge}$= 4.06 nm$^{-2}$s$^{-1}$.

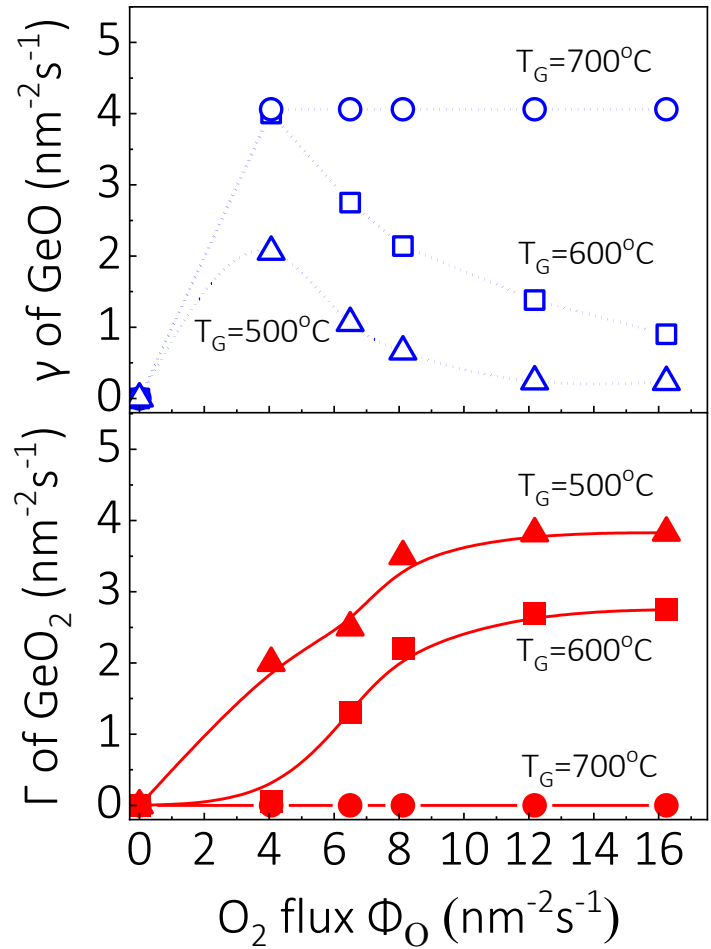

Figure 4. GeO$_2$ growth rate ($\Gamma$) measured by LR and GeO desorption rate ($\gamma$) simultaneously measured by QMS at different $T_G$ as a function of activated oxygen flux $\Phi_O$. The Ge flux is fixed at $\Phi_{Ge}$=4.06 nm$^{-2}$s$^{-1}$, corresponding to a growth rate of 1.7 Å/s at full Ge incorporation, i.e., for $\Gamma=\Phi_{Ge}$. The lines are guides to the eye.

The GeO$_2$ growth rates and corresponding GeO desorption rates for the layers of this sample are shown in Figure 4 as a function of $\Phi_O$ at different $T_G$. At $T_G$ of 500 °C and 600 °C, the growth rate increases with increasing $\Phi_O$ until it reached a plateau in the oxygen-rich regime. Under the oxygen-deficient condition, i.e., before the plateau, the given active oxygen flux is not sufficient to oxidize all the provided Ge flux to GeO$_2$. For instance, an oxygen flux of 4.06 O-atoms nm$^{-2}$s$^{-1}$ is equal to the supplied Ge flux, conforming to the stoichiometric ratio of the suboxide. The provided oxygen and Ge flux would thus form GeO. Depending on the specific growth temperature, the GeO could desorb (at ≥600°C in our case) or (partially) attach to the substrate (at 500°C in our case). The oxygen-deficiency-induced formation of the volatile suboxide can be evidenced by a decreasing GeO desorption rate as $\Phi_O$ increases.

In addition, it is apparent that the growth rate of GeO₂ is strongly temperature sensitive: At fixed $\Phi_O$ and $\Phi_{Ge}$, the growth rate decreases as $T_G$ increases from 500°C to 600°C and growth completely ceases at 700°C. This holds true even under oxygen-rich flux ratios and confirms the increased thermally activated GeO desorption with increasing $T_G$. For all growth conditions employed, the GeO-desorption equals the difference between the impinging Ge flux and the GeO₂ growth rate, i.e., any Ge that is not incorporated into the GeO₂ film desorbs as GeO.

In a previous study[32], a two-step reaction mechanism was established to describe the MBE growth of binary oxides like In₂O₃, Ga₂O₃ and SnO₂, in which the metal (e.g., Ga, In, Sn) oxidized to form a suboxide (e.g., Ga₂O, In₂O, SnO) via a first oxidization step and the suboxide further oxidized to form the solid metal oxide via a second oxidization step. Analogously, we applied this model to the growth of GeO₂. In this case, the first oxidization step is:

Ge(g) + O (g) → GeO(s).

The second oxidation step is:

GeO(s) → GeO₂(s)

and competes with the thermally activated suboxide desorption

GeO(s) → GeO(g).

Since GeO₂ and SnO₂ as well as their suboxides exhibit the same stoichiometry, we used the same rate equation for the chemical reaction and desorption of SnO₂ for the growth of GeO₂:

$$\Gamma = \frac{1}{2}\left(\Upsilon + \Phi_O - \sqrt{\Upsilon^2 + 2\Upsilon\Phi_O + (2\Phi_{Ge} - \Phi_O)^2}\right) \quad (1)$$

in which $\Upsilon = \Phi_{Ge} - \Gamma$, representing the suboxide desorption rate, and $\Upsilon$ can be further expressed as (related factorization and simplification can be seen in the Ref.[32]):

$$\Upsilon(R, T_G) = \Upsilon_0 e^{-\left(\frac{E_a(R)}{K_B T_G}\right)} \quad (2)$$

here, $\Upsilon_0 = e^{37.2}$, which is obtained by fitting the related vapor pressure data of the suboxide from literature and by converting the pressure to flux unit[38], $R$ is the ratio of $\Phi_{Ge}/\Phi_O$, $K_B$ is Boltzmann constant, the activation energy value $E_a$ is the only free parameter. Performing least-square fit of Eq. (1) to the $\Gamma$ of GeO₂ as a function of $T_G$ (Figure 5(b)), we are able to obtain $E_a$ as shown in Fig. S4, the value is linearly dependent on R:

$$E_a(R) = E_0 - \zeta R \quad (3)$$

By applying Eq. (3) and Eq. (2) to Eq. (1), Γ of GeO₂ can be modeled quantitatively as a function of $T_G$ and $\Phi_O$. The resulting model curves presented in Figure 5, are in good agreement withour measured growth rates as a function of $T_G$ and $\Phi_O$. The two-step oxidation model, thus, physically and quantitatively describes the PAMBE growth kinetics of GeO₂.

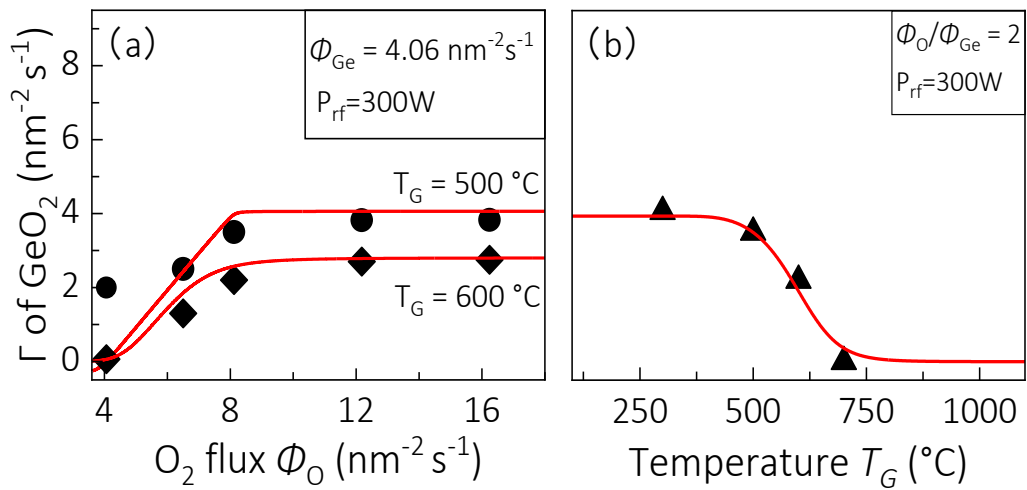

*Figure 5. Growth rate of GeO$_2$ as a function of (a) oxygen flux Φ$_O$ and (b) T$_G$. Black symbols represent measured Γ and red lines represent the model prediction.*

In order to predict the GeO$_2$-growth window, we used the model to calculate Γ at the fixed $Φ_{Ge}$=4.06 nm$^{-2}$s$^{-1}$ used in this study but for a wider range of $Φ_O$ and T$_G$ --- the results are shown in Fig. S5 and Fig. S6. Based on these calculated results we extract the predicted regions of different Ge-incorporation (>90% "full incorporation", 90%--10% "partial incorporation", <10% "no growth") shown in the $Φ_O/Φ_{Ge}$-T$_G$ space in Figure 6.

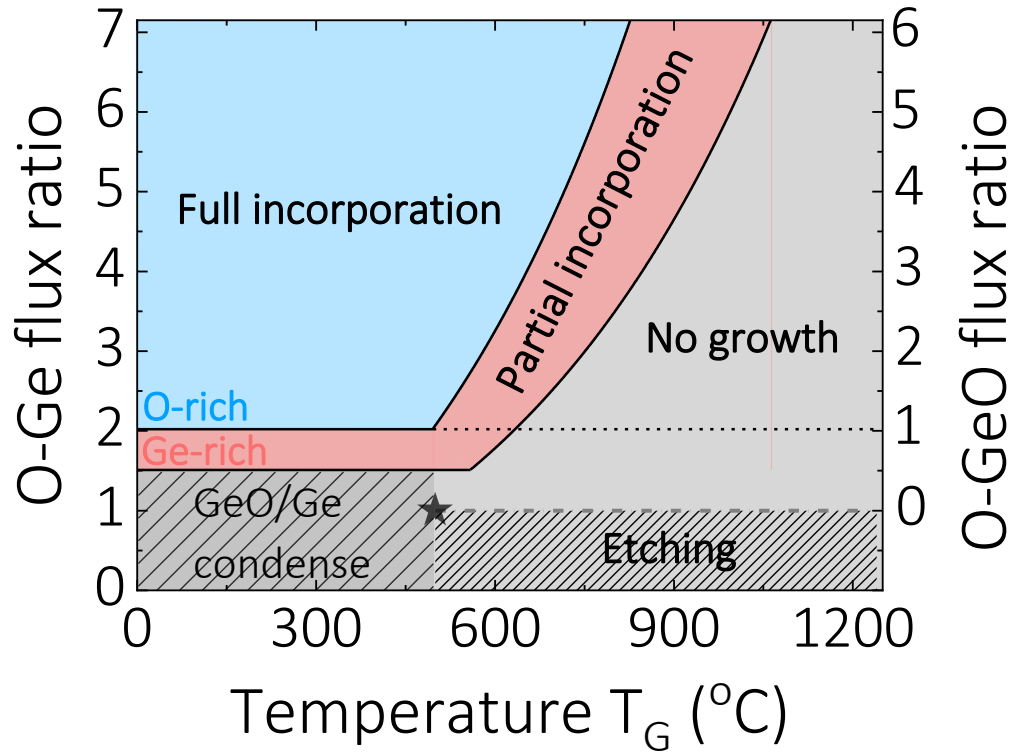

Figure 6. The PAMBE growth diagram of GeO$_2$ illustrating the growth window as a function of O-to-Ge flux ratio and $T_G$ modelled for the fixed Ge-flux of 4.06 nm$^{-2}$s$^{-1}$ used in our growth experiments. Horizontal lines separate the flux stoichiometry regions that enable O-rich growth, Ge-rich growth, and layer etching by Ge as labeled. The region of predicted condensation of secondary GeO phases into the film is marked by a hatched pattern and includes a star-shaped data point that corresponds to a condition in which we found condensation of GeO in our experiments. This growth diagram can also apply to the growth of GeO$_2$ by using GeO cell, which bypasses the intermediate formation of suboxide on the substrate, a case can be seen in ref [24].

The predicted GeO$_2$-growth window thus comprises the regions of full (blue shaded) and partial (red shaded) incorporation, whereas the grey-shaded region denotes the no-growth region. Full Ge-incorporation into GeO$_2$ can be found only in the $\Phi_O/\Phi_{Ge} \geq 2$-region (O-rich flux stoichiometry) at sufficiently low $T_G$ that prevents desorption of the intermediately-formed GeO. At a Ge-rich flux stoichiometry ($1 < \Phi_O/\Phi_{Ge} < 2$) the growth of pure GeO$_2$ is only possible at $T_G$ low enough (≤637°C in our example) to prevent full GeO desorption but high enough to prevent condensation secondary GeO-phases into the film. Impeded GeO desorption can, for example, explain the finite growth rate (of GeO) in Figure 4 and Figure 5(a) at $\Phi_O/\Phi_{Ge}=1$ ($\Phi_O=4.06$ nm$^{-2}$s$^{-1}$) and $T_G=500$°C -- these conditions are marked by the star-shaped datapoint in Figure 6.

Finally, at $\Phi_O/\Phi_{Ge}<1$ the fraction of provided Ge not oxidized into GeO can etch the GeO$_2$ film.

Lower(higher) $\Phi_{Ge}$ would shift the non-horizontal boundaries Figure 6 to lower(higher) temperatures to account for the decreased(increased) absolute GeO desorption. Since GeO desorption and oxidation into GeO$_2$ are also the main processes in suboxide MBE (providing a GeO instead of Ge flux) of GeO$_2$,[24] we conjecture our growth diagram to cover this growth method as well when offsetting the flux-ratio scale to account for absent consumption of O for the GeO formation. This scenario is described by the right-hand-side scale in Figure 6 and precludes layer etching, since this would require supply of an elemental Ge flux.

**Summary and Conclusion**

We have investigated the Ge incorporation and the competing desorption of its suboxide GeO during the growth of amorphous GeO$_2$ by plasma-assisted MBE from an elemental Ge source. The growth rate and desorbing flux were measured *in-situ* by laser reflectometry and line-of-sight quadrupole mass spectroscopy, respectively. For a detailed understanding of the reaction kinetics of GeO$_2$, the growth rate dependence on growth temperatures and oxygen fluxes was studied systematically. A Ge flux in the absence of oxygen lead to a negative growth rate, i.e., etching of the GeO$_2$ film. We generally demonstrated, that growth rates lower than the provided Ge flux were caused by GeO desorption driven by oxygen-deficiency or high temperatures. This behaviour was physically and quantitatively explained by a two-step oxidization model via the intermediate formation of GeO in good quantitative agreement with the measured growth rate as a function of all the studied growth parameters. Based on our results, we established a diagram describing the growth window and different growth regimes in parameter space spanned by O-Ge flux ratio and growth temperature.

We conclude, that our growth-kinetics study and the resulting growth diagram (Figure 6) can serve as a guidance for the thin film synthesis of all phases of GeO$_2$, since the oxidation reactions and GeO desorption are phase-independent. Potentially different surface free energies of crystalline GeO$_2$ phases, would be reflected by a different activation energy $E_a$ for GeO desorption and correspondingly re-scaled absolute temperatures in Figure 6.

We further conjecture our findings to also apply to the growth of GeO$_2$ by ozone- or suboxide-MBE. For all these flavors of MBE, O-rich growth conditions are recommended to (a) prevent condensation of secondary GeO phases into the film at low substrate temperature, and (b) prevent excessive GeO desorption at high substrate temperature. Ge-rich condition, on the other hand, may impact surface faceting (as observed during the MBE of In$_2$O$_3$)[39] or Ge/O-vacancy formation. Stabilization of crystalline phases of GeO$_2$ should be enabled by the choice of substrate or buffer layer (e.g., rutile-TiO$_2$ or rutile-SnO$_2$ for the growth of r-GeO$_2$)[40,41] likely within a limited, not too O-rich region[41] of the growth window.

Finally, the *in-situ* etching of GeO$_2$ by Ge, at elevated temperatures that enable GeO desorption, provides the potential for damage-free mesa etching in future GeO$_2$-device formation, a process already demonstrated for the case of Ga$_2$O$_3$ etched by Ga.[42]

**Acknowledgments**

The authors thank Elias Kluth, Martin Feneberg and Rüdiger Goldhahn for ellipsometric analysis of the GeO$_2$ stack. We further thank Hans-Peter Schönherr, Carsten Stemmler, and Steffen Behnke for technical support, as well as Johanna Nordlander for critically reading the manuscript. This work was performed in the framework of GraFOx, a Leibniz-ScienceCampus partially funded by the Leibniz association. W.C. gratefully acknowledges financial support from the Leibniz association under Grant No. K417/2021.

# In-situ study and modeling of the reaction kinetics during molecular beam epitaxy of GeO₂ and its etching by Ge

Wenshan Chen, Kingsley Egbo, Hans Tornatzky, Manfred Ramsteiner, Markus R. Wagner, Oliver Bierwagen

*Paul-Drude-Institut für Festkörperelektronik, Leibniz-Institut im Forschungsverbund Berlin e.V., Hausvogteiplatz 5–7, 10117 Berlin, Germany*

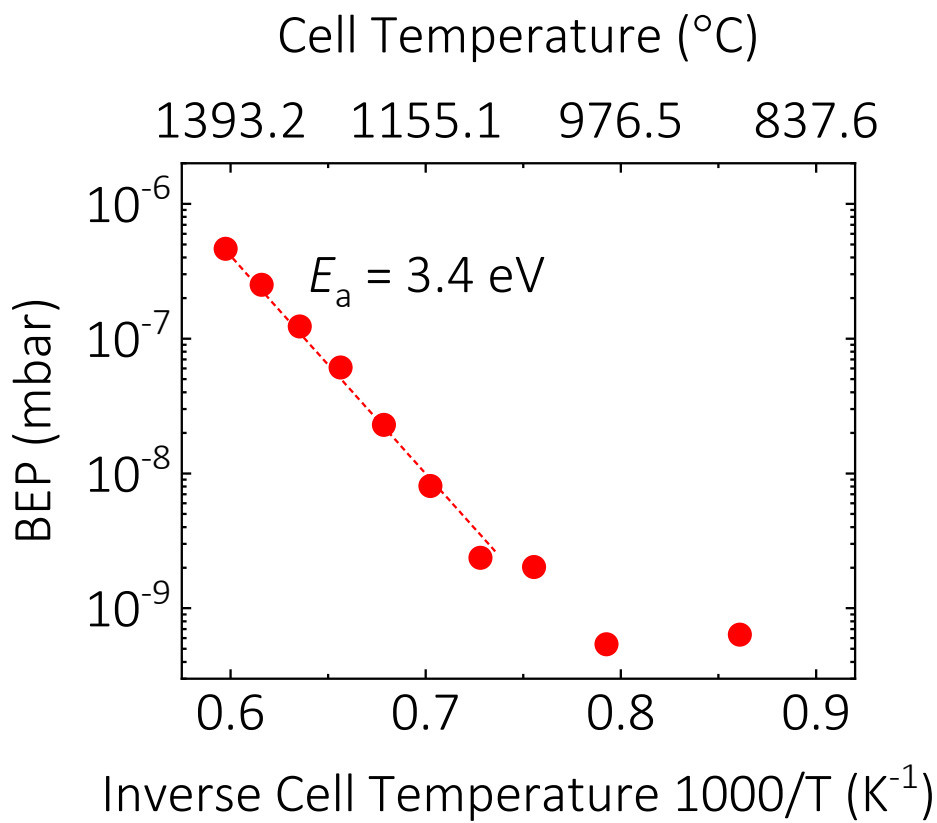

Figure S1. The beam equivalent pressure (BEP) as a function of inverse Ge-cell temperature. The symbols are measured values while the red dash line is the fitting line based on Arrhenius equation.

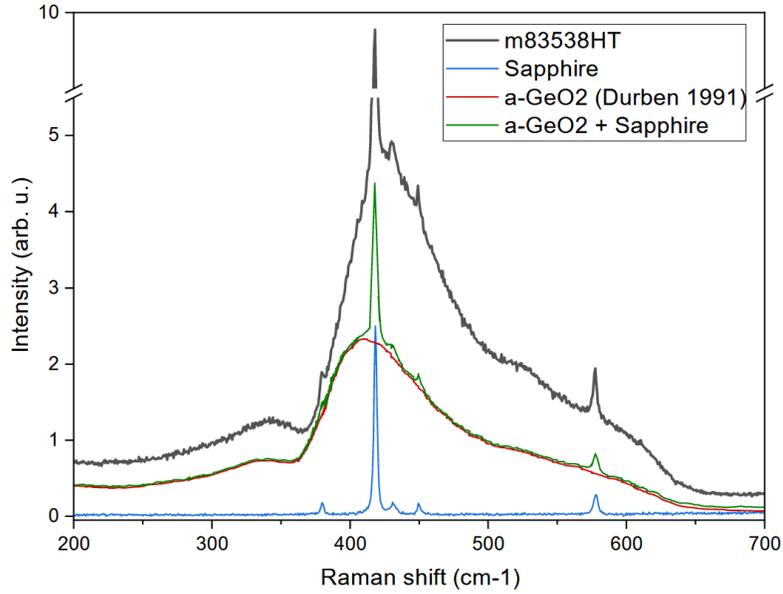

Figure S2. Corresponding Raman spectrum of the $GeO_2$ film described in Figure 3 as well as reference spectra measured for a pure sapphire substrate and amorphous $GeO_2$ from Ref. [16]. Excellent agreement of the broad feature can be found with the reference spectrum of a-$GeO_2$ while the observed sharp peaks are originated from the sapphire substrate.

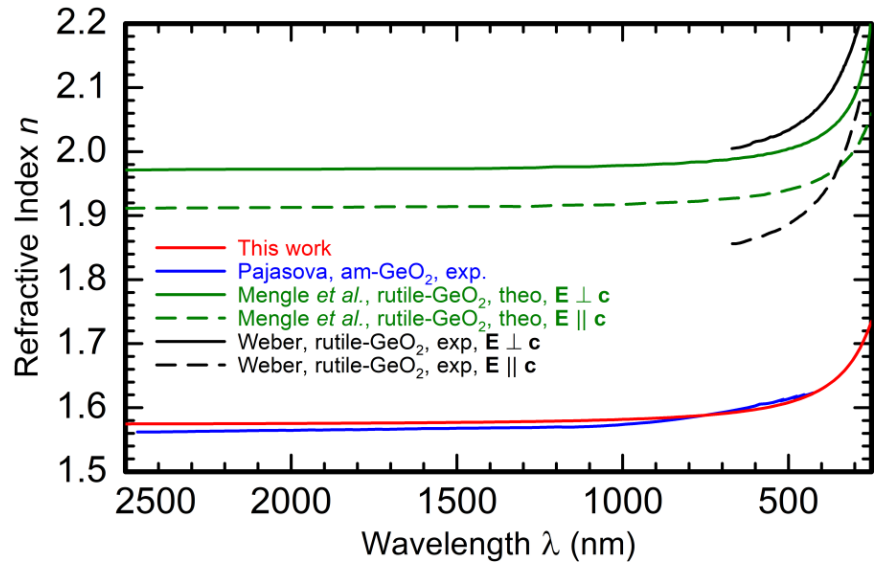

Figure S3. Average refractive index of the $GeO_2$ stack described in Figure 3 based on a Cauchy model (red) in comparison with experimental results of amorphous $GeO_2$ by Pajasova[43] (blue), of rutile $GeO_2$ by Weber[44] (black), and theoretical calculations of rutile $GeO_2$ by Mengle et al. [12] (green). In case of rutile $GeO_2$, both the ordinary (E ⊥ c) (solid) and the extraordinary component (E||c) (dashed) are displayed. The extracted extinction coefficient k of the $GeO_2$ stack was zero in the whole measurement range.

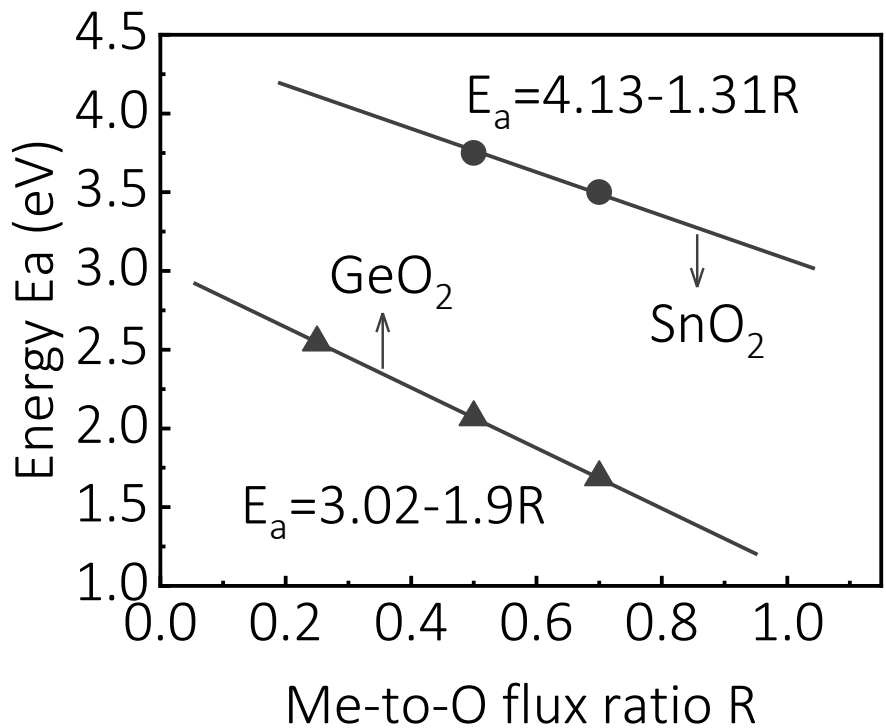

Figure S4. Activation energy $E_a$ for SnO and GeO desorption as a function of Me-to-O flux ratio. Symbols for $GeO_2$ are the data obtained by fitting the $T_G$ dependent $\Gamma$ in Fig 3(b), solid line is the linear fit of Eq. (3) to the data. The data for $SnO_2$ growth are taken from [32].

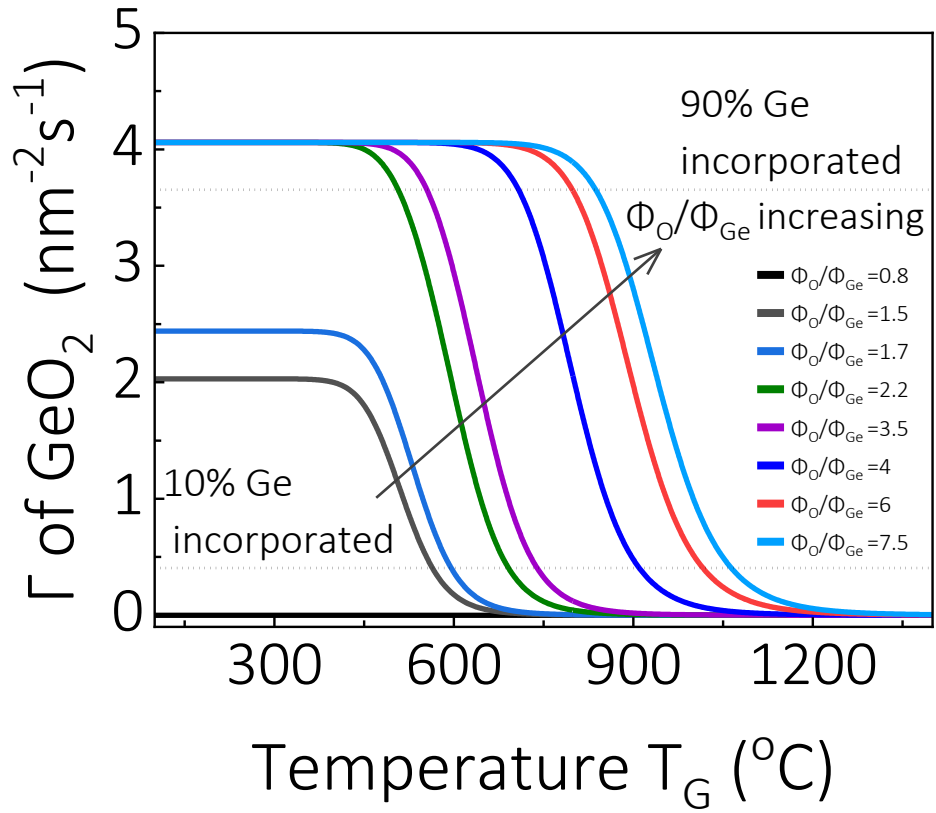

*Figure S5. Model predicted growth rate of GeO₂ as a function of T_G. The model paremeters are shown in Fig. S4.*

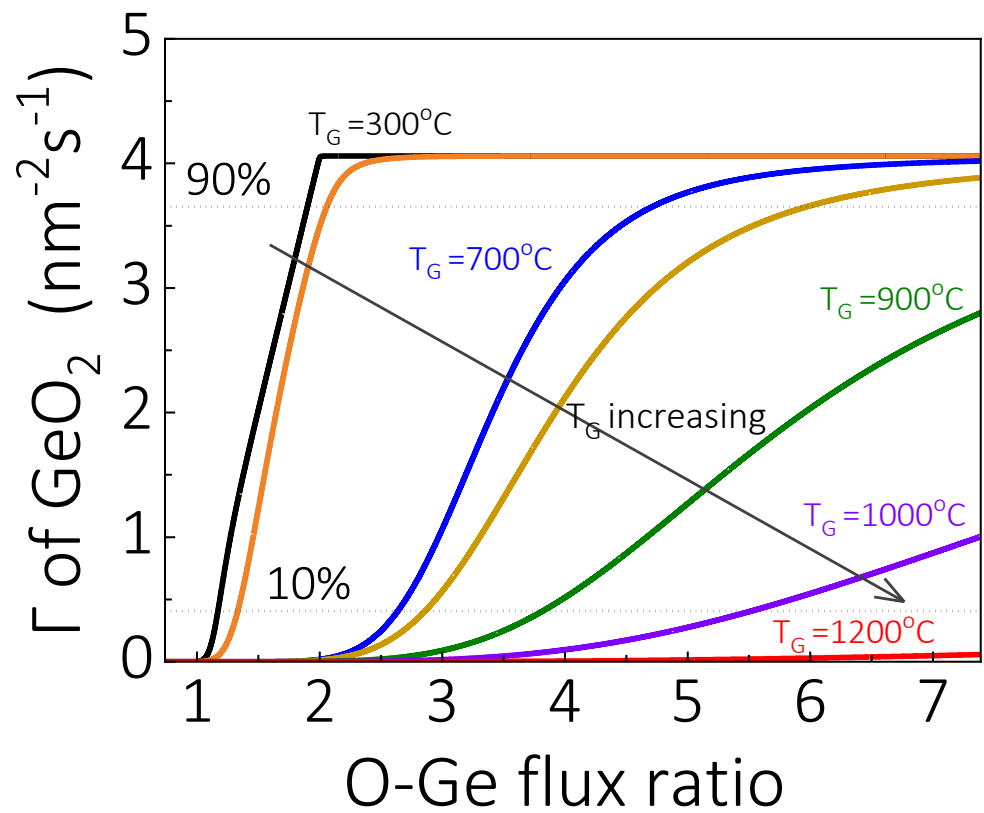

Figure S6. Model predicted growth rate of GeO₂ as a function of O-Ge flux ratio. The model paremeters are shown in Fig. S4.